\documentclass{article}
\usepackage{amssymb}

\usepackage{graphicx}
\usepackage{epsfig}

\begin{document}

\title{Multiple hits in wire chambers and other particle detectors.}

\author{Patrick Van Esch \\
Institut Laue Langevin \\
Grenoble - France}

\maketitle

\begin{abstract}
We propose an analysis of the dead time losses in counting imaging
detectors such as MWPC which can resolve $k$ simultaneous hits,
and analyze in more detail an $X-Y$ detector which has a third
wire set which allows for the recognition of simultaneous impacts.
\end{abstract}

\section{Introduction: the standard $X-Y$ detector.}

A particle detector with an intrinsic $X-Y$ structure, such as a
MWPC (see for example \cite{saulimwpc}) can be used as a pulse
counting 2-D imaging detector. A specific pixel of the imaging
detector shares a specific readout channel $x_1$ with other
pixels, and shares another specific readout channel $y_1$ with
other pixels, but in such a way that the combination of $x_1$ and
$y_1$ are not shared by any other pixel. A detector with $n_x
\times n_y$ pixels can hence be read out with $n_x + n_y$
channels.  As long as there is a single hit on the detector,
giving rise to a single $X$ and a single $Y$ channel being hit,
the hit pixel can be reconstructed without ambiguity. As long as
the signal development time of a single hit is small compared to
the time between impacts, a time correlation of $X$ and $Y$ hits
can associate the right $X$-hit with the right $Y$ hit, and within
this time window, a single hit will be able to be resolved into a
single pixel.

However, from the moment that two (or more) successive impacts on
the detector cannot be resolved in time, we have two (or more)
$X$-hits, say channels $x_1$ and $x_2$, and two (or more)
$Y$-hits, say channels $y_1$ and $y_2$, and it is not clear
whether the association should be $(x_1,y_1)$ and $(x_2,y_2)$ or
whether the association should be $(x_1,y_2)$ and $(x_2,y_1)$ ---
if there is no other information, such as amplitude correlations,
and the hits are similar. In other words, a double impact
generates 4 possible pixel candidates, of which 2 are the correct
ones, and 2 others are "ghost" hits. Let us consider that the
largest time which can separate the $X$ and the $Y$ signal of a
single hit, equals $\tau$, which we call the coincidence time. The
only way to associate unambiguously an $X$ hit to a $Y$ hit is
that during the time $\tau$ before and during the time $\tau$
after the event, no other event occurs. The probability in a
Poissonian stream with $n$ hits per second on average, that during
a time $T$ no event occurs, equals
\begin{equation}
\label{eq:probnoevent} P(T,k = 0) = \exp \left( - n T \right)
\end{equation}
The condition for an event, occurring at $t=0$ to be acceptable
(no ghosts), is that during the interval $t = -\tau$ to $t=0^-$,
no event happens, and that during the interval $t=0^+$ to
$t=\tau$, no event happens.  These probabilities being independent
for a Poisson stream (as the intervals are disjoint), the
probability for this to be so equals $P(\tau,k=0)^2 = \exp \left(
- n 2 \tau \right)$.  As such, the number of recorded hits (with
no ambiguity) in an $X-Y$ detector equals:
\begin{equation}
m = n \exp \left( - 2 n \tau \right)
\end{equation}
In other words, an $X-Y$ detector behaves as a paralyzable (see
for instance, \cite{knoll}) detector, but with \emph{twice} the
coincidence time $\tau$ as dead time.

Of course, there is also an intrinsic dead time for each $X$ wire
itself, and each $Y$ wire itself, which can often be longer than
the coincidence time window $\tau$.  If we call this wire dead
time, $\tau_w$, the time during which no two distinct hits on the
same $X$ wire can be distinguished, this will also give rise to a
dead time, in parallel with the coincidence dead time above.  In
order to estimate its importance, let us suppose that our detector
is uniformly irradiated.  This means that an $X$ wire has an
average Poissonian flux of $N_x = \frac{n}{n_x}$.  For a given
event at time $t=0$, we don't want its $X$ wire to be "blinded" by
an event preceding it, outside of the already considered
coincidence window.  The probability that no event occurs on the
same $X$ wire in the time window starting a time $t = -\tau_w$ but
stopping at $t = -\tau$, the start of the coincidence window,
equals, using equation \ref{eq:probnoevent}, $\exp \left( - N_x
(\tau_w-\tau) \right)$. We can apply the same reasoning for the
$Y$ wire, so the joint probability that no event occurs, nor on
the hit $X$ wire, nor on the hit $Y$ wire, so that both the $X$
and the $Y$ wire are not "blinded" before the actual coincidence
window opens, equals:
\begin{equation}
P_w = \exp \left( - \frac{n}{n_x} (\tau_w-\tau) \right) \exp
\left( - \frac{n}{n_y} (\tau_w-\tau) \right)
\end{equation}
The overall relationship between the true flux $n$, and the
observed flux $m$, in the case of a uniform, Poissonian
illumination, is:
\begin{equation}
m = n \exp \left( - 2 n \tau \right) \exp \left( - \frac{n}{n_x}
(\tau_w-\tau) \right) \exp \left( - \frac{n}{n_y} (\tau_w-\tau)
\right) = n \exp \left( - n \tau_t \right)
\end{equation}
where
\begin{equation}
\tau_t = \tau \left(2-\frac{1}{n_x} - \frac{1}{n_y}\right) +
\tau_w \left( \frac{1}{n_x}+\frac{1}{n_y}\right)
\end{equation}
As such, a standard $X-Y$ detector has a counting behaviour which
is that of a paralyzable detector with dead time given by
$\tau_t$.  We also see that this dead time consists of the
"coincidence dead time" ($2 \tau$) and has a contribution of the
"wire dead time" of the order of $2 \tau_w / n_x$.  So, the wire
dead time is less important (under uniform irradiation) when we
have $\tau_w/n_x \ll \tau$, which is often the case for large
MWPC.

\section{A third "disambiguation" electrode grid.}

An MWPC is often made by "sandwiching" an anode wire plane between
two cathode wire planes, or between a cathode wire plane and a
solid conductor surface.  One can make use of this degree of
freedom to use one cathode plane and the anode plane as the $X-Y$
channels, and the second cathode as a third set of channels.  In
the case of a wire plane, one can imagine for instance a set of
wires under 45 degrees with the $X-Y$ grid.  This has been
proposed by Lewis \cite{lewismwpc}.  It is also possible to use a
more symmetrical setup where the 3 wire planes (two cathode planes
and one anode plane in the case of a MWPC, or 3 different readout
directions on other types of detectors) make angles of 60 degrees
with each other, giving rise to some hexagonal 'honeycomb'
structure. This idea has been implemented using multi-GEM
detectors for X-ray imaging \cite{bachmannhex}, and the same idea
has also been proposed and experimented for Cherenkov photon
detectors (where there are typically simultaneous hits), in
\cite{saulihex}. In order to analyze the principle however, which
we set out to do in this paper, the exact geometry doesn't matter,
as long as the topological relationships between the coordinates
are the same.

Let us call this third set of channels, the $Z$ channels.  To each
possible hit corresponds, as before, an $x_i$ and a $y_j$, but now
also a $z_k$ channel hit, and the interesting point is that not
all combinations of $x$, $y$ and $z$ correspond to existing
pixels. In other words, for a single hit, there is some redundancy
in the information. This redundancy can be used to try to find out
what are the correct, and what are the "ghost" combinations of $x$
and $y$, when we have a multiple-hit event. One can easily
establish that it is always possible to find a numbering scheme of
the channels $X$, $Y$ and $Z$, such that, with a hit $(x,y)$,
there corresponds a $Z$-hit given by $z = x + y$.  It is herein
that lies the possibility to distinguish potentially multiple
hits: of a list of $X$ signals $\{x_1,x_2,...\}$, a list of $Y$
signals $\{y_1,y_2,...\}$ and a list of $Z$ signals
$\{z_1,z_2,...\}$, not all combinations are possible and one can
hope that only the correct combinations are allowed for.

\subsection{Two simultaneous hits.}

In fact, amongst two hits, the disambiguation is complete. Imagine
the hits $\{x_1,x_2\}$, $\{y_1,y_2\}$ and $\{z_1,z_2\}$ (where we
don't know of course the right numbering when we reconstruct
them). The right hits are $(x_1,y_1,z_1)$ and $(x_2,y_2,z_2)$.
This means that $z_1 = x_1 + y_1$ and $z_2 = x_2 + y_2$. The
question is: are there other possibilities ? Let us assume that
$x_1 \neq x_2$, $y_1 \neq y_2$ and $z_1 \neq z_2$ for starters.
Imagine that $(x_1,y_2,z_1)$ is a solution too. This means that
$z_1 = x_1 + y_2$, from which follows that $y_1 = y_2$ what was
against the starting hypothesis. A similar conclusion can be drawn
for the case $(x_1,y_2,z_2)$, and all other thinkable cases can be
reduced to these two by permutations.  On the other hand, imagine
that $x_1 = x_2$.  In that case, both the $(x_1,y_1,z_1)$ and
$(x_1,y_2,z_2)$ are the right solutions, and there are no others
(no ghosts can be constructed).

So this means that two simultaneous hits can be recognized
correctly with certainty. If we limit ourselves to this result,
then we can calculate the relation between true counting rate and
observed counting rate. An event will be lost, if there is more
than one (other) event in the time frame that starts time $\tau$
before and ends time $\tau$ after it. The Poisson distribution is
given by $P(k,\lambda) = \frac{\exp\left( - \lambda
\right)\lambda^k}{k!}$, from which it follows that $P(k <
u,\lambda) = \sum_{k=0}^{u-1} P(k,\lambda) =
\frac{\Gamma(u,\lambda)}{\Gamma(u)} = Q(u,\lambda)$ with $Q$ the
regularized incomplete gamma function (in \cite{abramowitz}, the
incomplete gamma function $P(a,x) = 1 - Q(a,x)$ is introduced ;
see also \cite{mathematica}). So the probability that an event
gets accepted is equal to $P(k<2,2 n \tau) = \Gamma(2,2 n \tau) =
e^{-2 n \tau}(1+2 n \tau)$, which is the probability that in the
interval of length $2 \tau$, centered onto the event under
discussion (but without counting this event of course) there is
one or no hits. As such, the observed counting rate will be:
\begin{equation}
\label{eq:doublehit}
 m = n e^{-2 n \tau}(1+2 n \tau)
\end{equation}
We see that (as far as coincidence dead time is concerned), a
detector that can discriminate up to two simultaneous hits, does
not exactly follow a paralysable or non-paralysable model, but
does in fact, much better: there is no first order dead time!
\begin{equation}
\label{eq:doublehitslow}
 m \simeq n - 2 \tau^2 n^3 + ...
\end{equation}

\subsection{$k$ simultaneous hits.}

\begin{figure}
  \includegraphics[width=12cm]{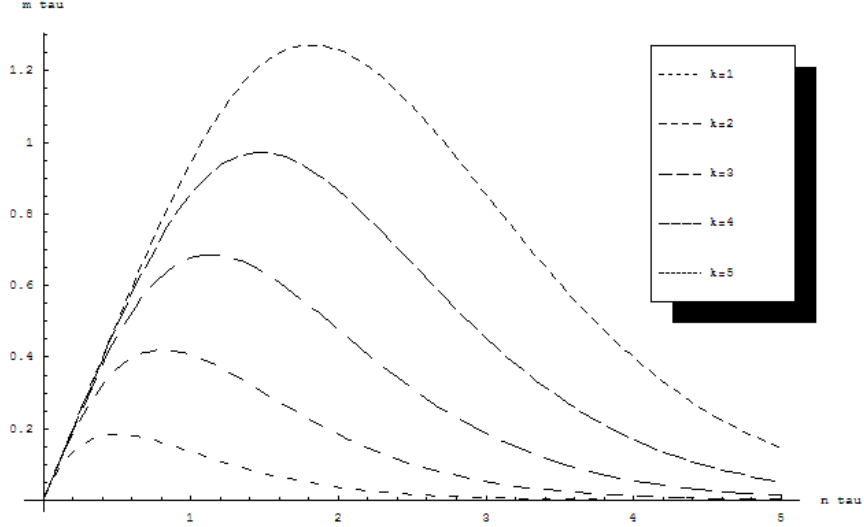}\\
  \caption{Observed counting rates as a function of incoming flux, when up to $k$ events can be
  discriminated in a time $\tau$.}\label{fig:deadtimeskhit}
\end{figure}

\begin{figure}
  \includegraphics[width=12cm]{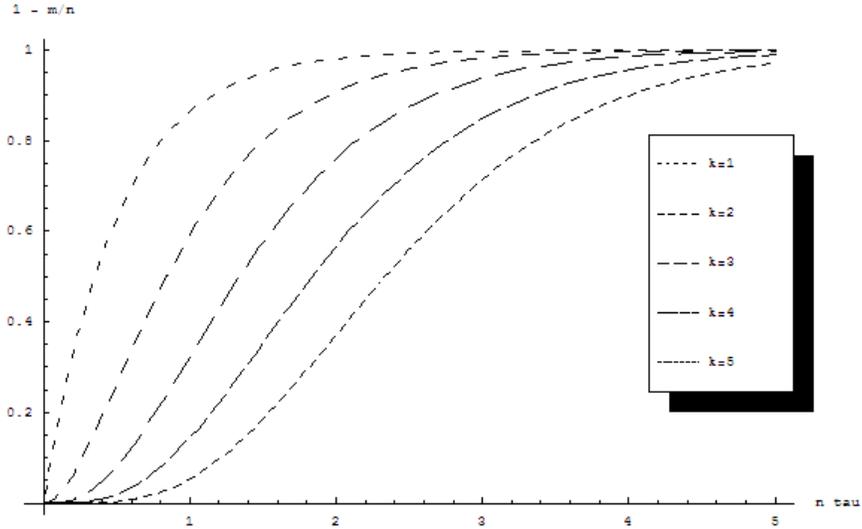}\\
  \caption{Dead time correction factor as a function of incoming flux, when up to $k$ events can be
  discriminated in a time $\tau$.}\label{fig:deadcorrectionkhit}
\end{figure}

Let us imagine for a moment that we have a detector that can
discriminate, without any difficulty, up to $k$ simultaneous
events.  What's the "dead time" now ? For a Poisson distribution,
in order for an event to be counted, we need to have less than $k$
events in an interval of $2\tau$ centered on an event to be
counted.  We will assume that the dead time is caused by the
coincidence time window needed, and is not dominated by individual
"wire" dead times. By a similar reasoning as above, we will obtain
that the observed counting rate will be:
\begin{equation}
m = n P(<k,2n\tau) = n Q(k,2 n \tau) = n
e^{-2n\tau}\sum_{u=0}^{k-1}\frac{(2 n \tau)^u}{u!}
\end{equation}
For $k=2$, we find the same expression as in equation
\ref{eq:doublehit}.  For $k=3$, we find:
\begin{equation}
m = n e^{-2n\tau}\left( 1+ 2n\tau + 2 n^2 \tau^2 \right)
\end{equation}
This is interesting, because we see that the last term neutralizes
the significant correction term in equation
\ref{eq:doublehitslow}.  We now have for low rates:
\begin{equation}
m \simeq n - \frac{4 \tau^3 n^4}{3}+...
\end{equation}
In figure \ref{fig:deadtimeskhit}, we show the relationship
between a (normalized on $1/\tau$) incoming Poissonian flux and
the (also normalized) counting rate of discriminated events for
$k=1,2,3,4,5$.  Often, the "dead-time correction" (which is given
by $1 - m/n$) shouldn't be too important if one wants to give
quantitative credibility to an imaging detector, and at the ILL,
we take as a definition of acceptable counting rate for a
detector, the one where the dead time correction is equal to 10\%.
The "dead time correction factor" $1-m/n$ is shown in figure
\ref{fig:deadcorrectionkhit}. In our case, we find then that this
counting rate $n_{10\%}$ is given by the solution of the following
equation:
\begin{equation}
Q(k,2 n_{10\%} \tau) = 0.9
\end{equation}
which leads to a solution using the inverse regularized gamma
function (see \cite{mathematica}):
\begin{equation}
n_{10\%} = \frac{Q^{-1}(k,0.9)}{2 \tau}
\end{equation}
Of course, the 10\% can be judged a bit too severe, and some
prefer allowing for 20\% correction, in which case we have:
\begin{equation}
n_{20\%} = \frac{Q^{-1}(k,0.8)}{2 \tau}
\end{equation}

Solving these equations for the first 5 $k$-values, we find:
\begin{center}
\begin{tabular}{||c|c|c||}
  \hline
  $k$ & $2 n_{10\%} \tau$  & $2 n_{20\%} \tau$ \\
  \hline
  \hline
  1 & $0.105$ & 0.223 \\
  \hline
  2 & $0.531$ & 0.824 \\
  \hline
  3 & 1.102 & 1.535 \\
  \hline
  4 & 1.745 & 2.297 \\
  \hline
  5 & 2.433 & 3.090 \\
  \hline
  \hline
\end{tabular}
\end{center}
So, for a traditional wire chamber, we find that a 10\% dead time
correction is reached when the flux is 0.105 times $1/(2 \tau)$;
when the chamber can distinguish 2 simultaneous hits, this flux is
5 times greater.  When the chamber can distinguish 3 simultaneous
hits, this flux is 2 times greater again, etc... If we adhere to
the definition of maximum flux at 20\% dead time correction, being
able to distinguish 2 simultaneous hits gives about 4 times higher
counting rates and being able to distinguish 3 simultaneous hits
adds a small factor 2 on top of this.

We hence see that the biggest gain in counting rate (as of our
definition using a maximum dead time correction of 10\% or 20\%)
occurs when we go from single-hit to double-hit identification,
which can, as shown previously, be obtained by using a third grid.

\subsection{How many hits can a third grid accept simultaneously ?}

If there are more than two simultaneous hits, then a third
electrode cannot guarantee that no ghost hit will be present.  Let
us consider the case of 3 simultaneous hits, and let us consider
that $(x_1,y_2,z_3)$ is a ghost hit.  If there is a ghost hit, we
can always permute indices to have the $X$ value be $x_1$ and the
$Y$ value be $y_2$.  In this case, the $Z$ value can only be
$z_3$, as any other value would bring us back to the case of two
hits and a ghost, which we demonstrated, is not possible.  So the
condition to have a ghost hit for a simultaneous impact of 3 hits
is:
\begin{equation}
x_1 + y_2 = x_3 + y_3
\end{equation}
(or a similar condition for permuted indices). Nevertheless, most
of the time when we have 3 hits the above condition will not be
satisfied. In that case, the signals still allow for a
non-ambiguous reconstruction of the 3 events. In order to find out
in detail what percentage of possible $k$-hit events can be
resolved one can take two roads: one is, for a specific setup, to
have a Monte-Carlo simulation of $k$-hit events ; the other is to
use some analytical estimations.  In any case, the percentage of
such resolved hits will depend also on the specific image that is
projected onto the detector: we will assume uniform irradiation in
our analysis.

\subsubsection{Analytical estimation of resolution of $k$-hit
events.}

Imagine that we have a $k$-hit event.  This means, a priori (we'll
come to that), that we have $k$ $X$ values, $k$ $Y$ values and $k$
$Z$ values.  We will assume (which is, depending on the exact
geometry of the detector, only approximately true), that each of
the $n_x$ $X$ values are equally probable, that each of the $n_y$
$Y$ values are equally probable, and that each of the $n_z$ $Z$
values are equally probable.  For each possible combination of an
$X$ value (in the hit list), and a $Y$ value (also in the hit
list), we can calculate the corresponding $Z$ value (using $z = x
+ y$). If this $Z$ value is in the list of $Z$ hits, we have to
accept the hit. Of course, for the real hits, this will be the
case.  It is for the $k(k-1)$ wrong combinations of $X$ and $Y$
values $(x_i,y_j)$ that we shouldn't, by coincidence, fall on a
$Z$ value $z_m$ which is also present in the hit list. By a
previous reasoning, we already know that this $z_m$ value cannot
be the $Z$ value which goes with the correct event of the $X$
value $x_i$, or of the $Y$ value $y_j$, so the potential index $m$
cannot be $i$ or $j$.  As such, there are $(k-2)$ possible $Z$
values which could, by coincidence, be equal to $x_i + y_j$.  In
order for our ghost couple $(x_i,y_j)$ to be rejected, none of
these should be equal to $x_i+y_j$. Assuming (which is an
approximation, but a reasonable one) that we can consider these
$(k-2)$ values (out of $n_z$ possible) as being statistically
independent, the probability for $x_i+y_j$ to be equal to one of
them equals then $\frac{k-2}{n_z}$, so the probability \emph{not}
to have a ghost hit for $(x_i,y_j)$ equals $1-\frac{k-2}{n_z}$.
There are $k(k-1)$ of these different combinations $(x_i,y_j)$
possible, so the probability (if, again, we consider all these
potential coincidences as statistically independent, which is of
course approximate but a reasonable hypothesis) that \emph{none}
of these combinations gives rise to a ghost hit (and hence, the
$k$-hit event is totally identifiable), is then:
\begin{equation}
\label{eq:pgood1} P_{good}^0 =
\left(1-\frac{k-2}{n_w}\right)^{k(k-1)}
\end{equation}
In the above formula, we have put $n_w = n_x = n_y = n_z$ the
number of channels, chosen to be equal for $X$, $Y$ and $Z$.  What
is clear is that when $n_w$ goes to infinity (infinitely many
channels per coordinate), that all $k$-fold hits are resolvable.
It is only due to a finite number of channels that 3 or more hits
can potentially give rise to ghost hits.

There is a caveat, however.  If we have $k$ hits and there is a
finite number of channels, then there is also a finite probability
that there are less than $k$ $X$-wires hit, or less than $k$ $Y$
wires hit or less than $k$ $Z$ wires hit because of the finite
probability to hit the same wire twice.

In the appendix in subsection \ref{subsec:hitcount}, the number
$Z^l_{k,n}$ is introduced, which gives us the number of different
ways one can construct an ordered list of $k$ elements, of which
each element is one of $n$ possible ones, and in which there are
exactly $l$ different elements present.  This corresponds to the
number of ways one can distribute $k$ hits over $n$ different
wires, and touch in all $l$ different wires.  Under a uniform
irradiation with "distinguishable" hits, each of these different
ways is equally probable, and hence the probability, if there are
$k$ hits, to have $l$ wires hit, is given by:
\begin{equation}
P(l;k,n_w) = \frac{Z_{k,n_w}^l}{n_w^k}
\end{equation}

So a better approximation for the probability of being able to
identify correctly a $k$-hit event, is to use formula
\ref{eq:pgood1} with the average value of $l$:
\begin{equation}
\langle l \rangle_k = \sum_{l=1}^k P(l ; k,n_w) l
\end{equation}
to give:
\begin{equation}
\label{eq:pgood2} P_{good}(k) = \left(1-\frac{\langle l \rangle_k
-2}{n_w}\right)^{\langle l \rangle_k(\langle l \rangle_k -1)}
\end{equation}

\subsubsection{Monte Carlo estimation of resolution of $k$-hits}

Let us consider a detector with 3 wire grids, at 120 degrees one
from the other and 32 wires each, equally spaced.  We delimit the
"useful" space as the hexagonal surface that is covered by the 3
different wire sets simultaneously, which consists of 768 pixels
(out of the 1024 crossing points of each pair of $32 \times 32$
wire planes). To each of these pixels correspond hence 3 numbers,
$x_i,y_i,z_i = x_i+y_i$.  A $k$-hit is formed by drawing
independently $k$ natural numbers between 1 and 768 (corresponding
to the pixels), and looking up what $x,y,z$ values they correspond
to, to make up the lists of $X$, $Y$ and $Z$ hits. Next, all
values of $(x_i + y_j)_{i\neq j}$ are tested against all values of
$z_u$ with $u \neq i,j$: these are the ghosts that cannot be
rejected ; we count how many there are. We repeat this procedure a
large number of times (number of trials $N$) and make then a
histogram of the number of ghosts that each trial generated.  Zero
ghosts means that the event could be correctly reconstructed. The
fraction of the number of zero ghosts $N_0(k)$ over $N$ is an
estimator of the probability to be able to reconstruct a $k$-hit
event.

For $N = 10000$, we find:
\begin{center}
\begin{tabular}{|c|c|c|c|c|c|c|c|c|c|c|c|c|}
  \hline
  $k$ & 1 & 2 & 3 & 4 & 5 & 6 & 7 & 8 & 9 & 10 & 11 & 12  \\
  \hline
  $N_0(k)$ & 10000 & 10000 & 8639 & 5772 & 2780 & 918 & 215 & 41 & 7 & 1 & 0 & 0  \\
  \hline
\end{tabular}
\end{center}

If we compare that with our analytical estimation in equation
\ref{eq:pgood2} (multiplied with $N$), then we obtain:
\begin{center}
\begin{tabular}{|c|c|c|c|c|c|c|c|c|c|c|c|c|}
  \hline
  $k$ & 1 & 2 & 3 & 4 & 5 & 6 & 7 & 8 & 9 & 10 & 11 & 12  \\
  \hline
  $N P_{good}(k)$ & 10000 & 10018 & 8526 & 5336 & 2167 & 513 & 65 & 4 & 0 & 0 & 0 & 0  \\
  \hline
\end{tabular}
\end{center}
which gives quite good agreement especially for the lower $k$
values.  For the higher $k$ values, there is an under-estimation
of the number of correctly identified $k$-hits.

Repeating the experience with 128 wires per plane, we find, after
10000 trial events, for the Monte Carlo result:
\begin{center}
\begin{tabular}{|c|c|c|c|c|c|c|c|c|c|c|c|c|}
  \hline
  $k$ & 1 & 2 & 3 & 4 & 5 & 6 & 7 & 8 & 9 & 10 & 11 & 12  \\
  \hline
  $N_0(k)$ & 10000 & 10000 & 9612 & 8605 & 6820 & 4750 & 2701 & 1326 & 461 & 153 & 24 & 8  \\
  \hline
\end{tabular}
\end{center}
while the analytical estimation gives us:
\begin{center}
\begin{tabular}{|c|c|c|c|c|c|c|c|c|c|c|c|c|}
  \hline
  $k$ & 1 & 2 & 3 & 4 & 5 & 6 & 7 & 8 & 9 & 10 & 11 & 12  \\
  \hline
  $N P_{good}(k)$ & 10000 & 10001 & 9559  & 8356 & 6402 & 4123 & 2148 & 869 & 264 & 58 & 9 & 1  \\
  \hline
\end{tabular}
\end{center}

Again, one observes a relatively good performance for the lower
$k$-values, and an underestimation of the number of correctly
identified events at higher $k$ values.

\subsection{Count rates, dead times, and the third coordinate.}

\begin{figure}
  \includegraphics[width=12cm]{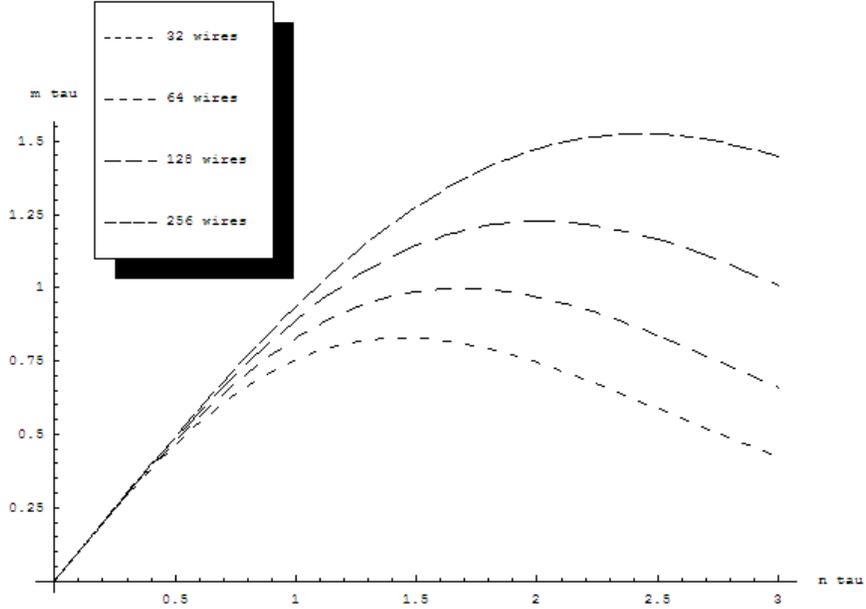}\\
  \caption{The observed (normalized) counting rate of identified hits as a function
  of the (normalized) incoming flux, using equation \ref{eq:deadtimesnwires}.}\label{fig:deadtimesnwires}
\end{figure}

\begin{figure}
  \includegraphics[width=12cm]{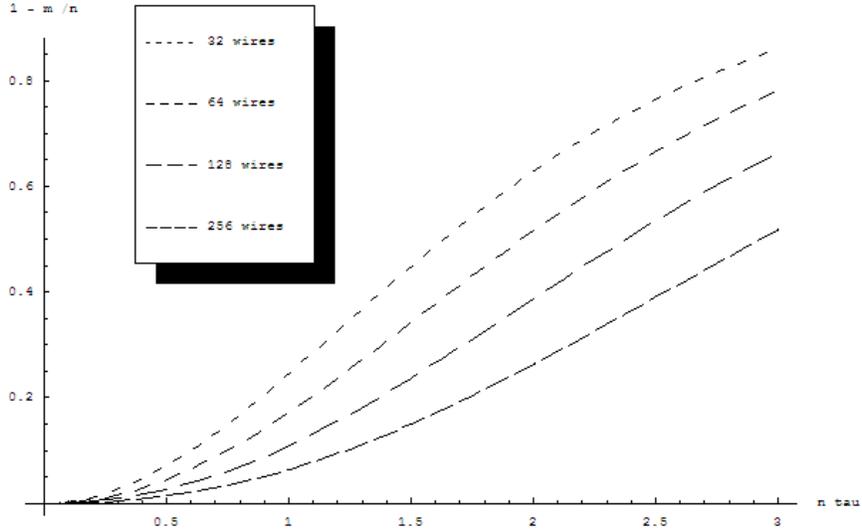}\\
  \caption{The dead time correction of identified hits as a function
  of the (normalized) incoming flux, using equation \ref{eq:deadtimesnwires}.}
  \label{fig:deadcorrectionnwires}
\end{figure}

Previously, we considered detectors which could, with 100\%
certainty, discriminate an event when there were no more than
$k-1$ other events in a time slot $2 \tau$ centered on our event,
and which couldn't handle, also with certainty, an event when
there were $k$ or more events in the given time slot. The detector
with a third coordinate however, has a different behavior: it can
discriminate, with a certain probability $P_{good}(k)$, the case
where there are $(k-1)$ events in the said time  slot (on top of
our event-under-test). This means that the relationship between
incoming flux and observed, identified rate of hits is now given
by:
\begin{equation}
\label{eq:deadtimesnwires} m = n \sum_{k=1}^{\infty}
P_{Pois}(k-1,2\tau n)P_{good}(k)
\end{equation}
a sum which we can of course truncate to the first few terms given
that as well the probability to have a $k$-hit as well as the
probability to resolve it correctly, will drop fast for high $k$
values.  As such, our analytical estimate can be used, given that
its performance for relatively low $k$ values is adequate.  Using
equation \ref{eq:deadtimesnwires}, we can plot (figure
\ref{fig:deadtimesnwires}) the observed (normalized) counting rate
of identified events as a function of the incoming flux, for
different numbers of wires per plane.  The associated dead time
correction is shown in figure \ref{fig:deadcorrectionnwires}.
Calculating the incoming fluxes that give rise to a dead time
correction of 10\% or 20\%, we find (normalized onto $2 \tau$ as
was the case in the fixed $k$ case):
\begin{center}
\begin{tabular}{||c|c|c||}
  \hline
  Number of wires per plane & $ n_{10\%} 2 \tau$  & $ n_{20\%}  2 \tau$ \\
  \hline
  \hline
  32 & 1.203 & 1.771 \\
  \hline
  64 & 1.502 & 2.178 \\
  \hline
  128 & 1.917 & 2.734 \\
  \hline
  256 & 2.472 & 3.467 \\
  \hline
  \hline
\end{tabular}
\end{center}

\section{Conclusion.}

In this paper we discussed the theoretical dead time correction
occurring in detectors which need a certain time to recognize the
correct positioning of an impact in the case that they can handle
$k$ impacts simultaneously. We applied this to the specific case
of a detector with an $X-Y-Z$ planar wire structure.  It is
established that in principle, such a detector (which has 50\%
more readout channels than a standard $X-Y$ detector) can accept
incoming counting rates which are about 12 times (or more) larger
than the standard $X-Y$ detector at 10\% dead time correction.  As
such, the investment in 50\% more channels can result potentially
in 1200\% higher counting rates.  However, these results are only
established in the case of a Poissonian, uniform irradiation.

We established a relatively simple analytical estimate of the dead
time correction, and verified its applicability with a more
detailed Monte Carlo simulation.

\newpage

\appendix

\section{Counting multi-hits.}

\subsection{Counting distinguishable hits.}
\label{subsec:hitcount}

Let us define a $k$-fold multiple hit event on a set of "pixels"
with $n$ possible "positions" as an ordered series
$\{x_1,x_2,...,x_k\}$, where $x_i$ takes its values out of the set
$\{1,2,...n\}$.  The point is of course that the $j$-th hit can be
identical to the $i$-th hit, so this is the well-known problem of
an \emph{ordered set with repetitions}.  The number of different
such $k$-fold hit events is then given by the formula:
\begin{equation}
Z_{k,n} = n^k
\end{equation}
where we introduced the symbol $Z_{k,n}$ to stand for the number
of ordered sets of $k$ hits out of, each time, $n$ possibilities.
Let us remind, for completeness, that, if we were not to allow
repetitions (that is, we require the $k$ hits to be different),
that the number of possibilities is the number of
\emph{permutations}:
\begin{equation}
P^n_k = \frac{n!}{(n-k)!}
\end{equation}

We now want to find out, in the case of ordered sets with
repetitions, what is the number of different $k$-fold hit events,
in which there are exactly $l$ different values occurring in the
series, where of course $l \leq k$ and $l \leq n$.  Indeed,
because of the possibility of repetitions, some of the $k$
different drawings can be identical, and we want to know how many
different cells are finally hit by the $k$ hits.  We will call
this number: $Z_{k,n}^l$. We didn't find any explicit reference to
this number in the literature, but one can easily find a recursive
definition for it.  First, we consider special cases:
\begin{itemize}
    \item The case $l = k$.  We want all $k$ hit cells to be
    different.  This is just drawing without repetition:
    \begin{equation}
    Z_{k,n}^k = P^n_k = \frac{n!}{(n-k)!}
    \end{equation}
    \item The case $l = 1$.  All $k$ hits are identical.  Clearly,
    there are exactly $n$ different ways to do so:
    \begin{equation}
    Z_{k,n}^1 = n
    \end{equation}
    \item The recurrence step.  In order to compute $Z_{k,n}^l$,
    we consider that we have already $(k-1)$ hits, and we add the
    last hit.  There are two possibilities: or there are already
    $l$ different elements hit in the $(k-1)$ first hits, in which
    case the $k$-th hit must be one of these $l$ elements ; or
    there were only $(l-1)$ different elements hit in the first
    $(k-1)$ hits, and hence the last hit must be a different hit,
    drawn from the $(n-l+1)$ remaining un-hit cells.  As such, we
    obtain:
    \begin{equation}
    Z_{k,n}^l = Z_{k-1,n}^l.l + Z_{k-1,n}^{l-1}.(n-l+1)
    \end{equation}
\end{itemize}
The two special cases and the recurrence relation define the
function $Z_{k,n}^l$ completely: indeed, at each recurrence step,
$k$ is diminished by 1, while $l$ remains, or is diminished by 1.
So sooner or later, $k$ will reach the value of $l$, in which case
the first special case is applicable, or $l$ will reach the value
of $1$, in which case the second special case is applicable.

One should obtain that:
\begin{equation}
\sum_{l=1}^k Z_{k,n}^l = Z_{k,n} = n^k
\end{equation}

Indeed, when summing over all possibilities of exactly $l$
different cells, we obtain of course all the possibilities of
drawing $k$ hits out of $n$ possible cells.  We didn't succeed in
obtaining an entirely closed form for the solution, but we did
find solutions in closed form for small values of $l$
\footnote{These relations are found iteratively by solving the
single-variable recursion relation in $k$ for respectively $l=2$,
$l=3$, $l=4$ ; however, in order to write out this recursion
relation for $l$, we need already the explicit expressions for
$l-1$.  For instance, for $l=2$, the recursion relation is found
to be: $Z_{k,n}^2 = n(n-1) + 2 Z_{k-1,n}^2$, which is a pure
recursion relation over $k$.  We can solve it by standard
difference equation techniques.  Once we know $Z_{k,n}^2$, we can
write a recursion relation for $Z_{k,n}^3 =
(n-2)\underline{Z_{k-1,n}^2} + 3 Z_{k-1,n}^3$ in which the
underlined part is replaced by its (now known) explicit
expression, so that only an explicit recursion in $k$ remains.  In
all generality, we have: $Z_{k,n}^l =
(n-l+1)\underline{Z_{k-1,n}^{l-1}} + l Z_{k-1,n}^l$, again, with
the underlined part replaced by its explicit closed-form
expression.}:
\begin{eqnarray}
  Z_{k,n}^2 &=& \frac{\left( -2 + 2^k \right) \,\left( -1 + n \right) \,n}{2} \\
  Z_{k,n}^3 &=& \frac{\left( 27 - 27\,2^{1 + k} + 17\,3^k \right) \,
    \left( -2 + n \right) \,\left( -1 + n \right) \,n}{54} \\
  Z_{k,n}^4 &=& \frac{\left( -384 + 9\,2^{8 + k} - 2176\,3^k + 555\,4^k \right)
    \left(n-3 \right)\left( n-2 \right)
    \left( n-1 \right)n}{2304}
\end{eqnarray}

\subsection{Counting indistinguishable hits.}

Consider, for completeness\footnote{In usual wire chambers, the
quantum-mechanical indistinguishability of multi-particle states
doesn't play a role as the phase-space resolution given by the
"pixelisation" in space and time of the detector is several orders
of magnitude more coarse than the quantum-mechanical
"pixelisation".  Nevertheless, in the case where the bosonic
character of the particle states plays a role, we only have to
replace $Z$ by $K$.}, the case where the hits are
indistinguishable in principle. In that case, the order of the
hits in a $k$-hit event is unimportant.  We can introduce a
similar quantity to $Z_{k,n}^l$, namely, $K_{k,n}^l$, with the
meaning of $l$ different cells being hit in a $k$-hit event out of
$n$ possible different cells, but this time where the order is
unimportant (though the multiplicity is). The total number of
different (unordered) multisets with repetition with cardinality
$k$ drawn out of $n$ possible values is known to be:
\begin{equation}
\langle
\begin{array}{c}
  n \\
  k \\
\end{array}%
\rangle
=
\left(%
\begin{array}{c}
  n+k-1 \\
  k \\
\end{array}%
\right) =
\left(%
\begin{array}{c}
  n+k-1 \\
  n-1 \\
\end{array}%
\right)
\end{equation}

Here, $\left(%
\begin{array}{c}
  n \\
  k \\
\end{array}%
\right) = C_n^k $ is the binomial coefficient.

If we consider $k$ unordered hits without repetition then we
simply have the number of combinations, given by the binomial
coefficient.  From this, we can deduce that:
\begin{equation}
K_{k,n}^k = C_n^k
\end{equation}
We can also deduce that:
\begin{equation}
\sum_{l=1}^k K_{k,n}^l = \langle
\begin{array}{c}
  n \\
  k \\
\end{array}%
\rangle
\end{equation}

In order to derive a general expression for $K_{k,n}^l$, we
consider the following ordering of the $k$ hits: the first $l$
cells are the cells which have distinct hits, and the last $k-l$
cells are the cells which have already been hit.  The first $l$
cells (all different) can be chosen in $C_n^l$ different ways out
of the $n$ existing cells.  The last $k-l$ cells have to be all
the unordered drawings with repetition out of the $l$ cells that
have already been chosen in the first part: there are $\langle
\begin{array}{c}
  l \\
  k-l \\
\end{array}%
\rangle$ ways of doing this.  So we conclude that:
\begin{equation}
K_{k,n}^l = C_n^l . C_{k-1}^{k-l}
\end{equation}
and we have an explicit, closed-form expression for the unordered
multisets of cardinality $k$, with exactly $l$ distinct elements,
drawn out of $n$ possibilities.

\section{Dead times.}

The classical models of paralyzable and non-paralysable detectors
(see \cite{knoll} for instance) are defined as follows: a
non-paralyzable detector is "blind" for a fixed time $\tau$ after
an event is registered (no matter what happens during this time),
and is sensitive again after this time.  The relationship between
the true rate $n$ and the observed rate $m$ is then given by:
\begin{equation}
n = \frac{m}{1-m \tau}
\end{equation}
If the detector is "far from saturation" (relatively few events
are lost), then we can write:
\begin{equation}
m \simeq n(1-\tau n)
\end{equation}
In the case of a paralyzable detector, the detector is "blind" for
at least a time $\tau$ after an event is registered, but each time
there is a new (uncounted) event during the "blind period", the
detector remains blind for a time $\tau$ after this event.  In
other words, the detector becomes ready again only if there is a
time of "silence" after an event of at least $\tau$.  A detectable
event must hence be preceded by a silence of at least $\tau$. We
summarise the derivation given in \cite{knoll}, because it will be
instructive for our own developments in this text. The
probability, given an event at time $t=0$, to have the next event
happening between time $t$ and time $t+dt$, equals, for a Poisson
stream:
\begin{equation}
I_1(t)dt = n \exp(-n t)dt
\end{equation}
The probability to have an event occurring after a "silence" of at
least time $\tau$ is then given by
\begin{equation}
\int_{t=\tau}^\infty I_1(t) dt = \exp(-n\tau)
\end{equation}
This gives then:
\begin{equation}
m = n \exp(-n \tau)
\end{equation}
If the detector is "far from saturation", then we can write also:
\begin{equation}
m \simeq n (1 - \tau n)
\end{equation}
so, far from saturation, there is no difference in behavior
between a paralyzable and a non-paralyzable detector.

\end{document}